\documentclass[preprint,12pt,authoryear]{elsarticle}
\pdfoutput=1

\usepackage{hyperref}
\hypersetup{colorlinks=true, linkcolor=red, citecolor=blue}
\usepackage{chngpage}
\usepackage{soul}
\usepackage{ulem}
\usepackage{color}
\usepackage{wasysym}
\usepackage{amssymb, amsthm}
\usepackage{graphicx}
\usepackage{float}
\usepackage[utf8]{inputenc}

\def\la{\raise.5ex\hbox{$<$}\kern-.8em\lower 1mm\hbox{$\sim$}}
\def\ga{\raise.5ex\hbox{$>$}\kern-.8em\lower 1mm\hbox{$\sim$}}
\def\be{\begin{equation}}
\def\ee{\end{equation}}
\def\ba{\begin{eqnarray}}
\def\ea{\end{eqnarray}}

\def\Omegastar{\Omega_\ast}

\def\Mdotin{\dot{M}_{\mathrm{in}}}
\def\Mdot{\dot{M}}
\def\Edot{\dot{E}}
\def\Pdot{\dot{P}}

\def\Msun{\mathrm{M}_{\astrosun}}

\def\Bd{B_{\mathrm{d}}}

\def\rlc{r_{\mathrm{LC}}}

\def\Lx{L_{\mathrm{x}}}

\def\Md{M_{\mathrm{d}}}

\def\rA{r_{\mathrm{A}}}

\def\Tp{T_{\mathrm{p}}}

\def\dM*{\delta M_*}

\def\Firr{F_{\mathrm{irr}}}

\def\P0min{P_{0,{\mathrm{min}}}}

\def\Alfven{Alfv$\acute{e}$n}
\def\418{SGR 0418+5729}
\def\142{AXP 0142+61}
\def\ql{\textquoteleft}
\def\qr{\textquoteright}

\def\rbb1{R_{\mathrm{BB1}}}
\def\rbb2{R_{\mathrm{BB2}}}

\journal{New Astronomy}

\begin{document}

\begin{frontmatter}



\title{Rotational and X-ray luminosity evolution of high-B radio pulsars }


\author{Onur Benli, \"{U}nal Ertan}

\address{Sabanc\i\ University, 34956, Orhanl\i\, Tuzla, \.Istanbul, Turkey}
\ead{onurbenli@sabanciuniv.edu}

\begin{abstract}

In continuation of our earlier work on the long-term evolution of the so-called high--B radio pulsars (HBRPs) with measured braking indices, we have investigated the long-term evolution of the remaining five HBRPs for which braking indices have not been measured yet.  This completes our source-by-source analyses of HBRPs in the fallback disc model that was also applied earlier to anomalous X-ray pulsars (AXPs), soft gamma repeaters (SGRs), and  dim isolated neutron stars (XDINs). Our results show that the X-ray luminosities and the rotational properties of these rather different neutron star populations can be acquired by neutron stars with fallback discs as a result of differences in their initial conditions, namely the initial disc mass, initial period and the dipole field strength. For the five HBRPs, unlike for AXPs, SGRs and XDINs, our results do not  constrain the dipole field strengths of the sources. We obtain evolutionary paths leading to the properties of HBRPs in the propeller phase with dipole fields sufficiently strong to produce pulsed radio emission.   

\end{abstract}

\begin{keyword}
accretion, accretion discs \sep magnetic fields \sep (stars:) pulsars: individual (HBRPs) \sep X-rays: stars


\end{keyword}

\end{frontmatter}


\section{Introduction}
\label{sec1}

Rotational evolution of an isolated pulsar in vacuum is governed by magnetic dipole torque. In this case, the magnetic dipole field strength on the pole of the neutron star can be estimated from the observed period, $P$, and period derivative, $\Pdot$, as $B \simeq 6.4 \times 10^{19} \sqrt{P \Pdot}$~G for a neutron star with moment of inertia $I = 10^{45}$~g~cm$^2$ and radius $R = 10^6$~cm. Throughout this work, we reserve \ql$\Bd$\qr~ to denote the field strength inferred from the dipole torque formula. For most normal radio pulsars $\Bd \sim 10^{12}$ G, while for the newly discovered neutron star populations, namely Anomalous X-ray pulsars (AXPs), soft gamma repeaters (SGRs), dim isolated neutron stars (XDINs), and high-B radio pulsars (HBRPs), $\Bd$ values are greater than the quantum critical limit, $B_{\mathrm{c}} \equiv m_{\mathrm{e}}^2 c^3 / e \hslash = 4.4 \times 10^{13}$~G. Rotational periods of XDINs, AXPs, and SGRs clustered to the $2 -12$ s range, and super-Eddington soft gamma bursts, which were once considered to be the distinguishing property of SGRs, emitted also by some of the AXPs \citep{Kaspi_etal_03, Israel_etal_07} and HBRPs \citep{Gavriil_etal_08, Younes_etal_16} seem to indicate evolutionary connections between these neutron star populations \citep{Keane_Kramer_08, Kaspi_10}.   

In the magnetar model \citep{Thompson_Duncan_95}, these young neutron star systems are assumed to slow down in vacuum, and the actual field strength, $B_0$, on the star is taken to be equal to $\Bd$. If these systems are evolving with fallback discs estimated to have been formed after the supernova \citep{Michel_Dessler_81, Michel_88, Chevalier_89}, as proposed by \cite{Chatterjee_etal_00, Alpar_01}, the $B_0$ values are estimated to be at least an order of magnitude smaller than $\Bd$. Because the disc torques usually dominates the dipole torques during the evolution of a neutron star with a fallback disc. 

Recently, it was shown that the long-term X-ray luminosity and the rotational evolution of AXP/SGRs, XDINs, and HBRPs with measured braking indices can be explained in the same fallback disc model \citep{Ertan_etal_14, Benli_Ertan_16, Benli_Ertan_17}. In the present work, we investigate the evolution of the remaining HBRPs for which X-ray luminosities, periods and period derivatives are known but braking indices have not been measured yet. In Section \ref{sec2}, we briefly describe the model parameters. Observational properties of HBRPs that we investigate in this work are summarised in Section \ref{sec3}. We discuss the model results, and summarise our conclusions in Section \ref{sec4}.

\section{Model Parameters}
\label{sec2}

A detailed description of our long-term evolution model is given in \cite{Ertan_etal_14, Benli_Ertan_16} and references there. In this section, we briefly describe the model parameters. There are three basic disc parameters that are expected to be similar for the fallback discs of different systems: the X-ray irradiation efficiency ($C$), the minimum critical temperature below which the disc becomes viscously passive ($\Tp$),  and the $\alpha$ parameter of the kinematic viscosity \citep{Shakura_Sunyaev_73}. In this work, we use the same values of the $C$, $\Tp$, and $\alpha$ parameters as those employed earlier in the evolution models of individual AXP/SGRs \citep{Benli_Ertan_16}, XDINs \citep{Ertan_etal_14}, and the three HBRPs with measured braking indices \citep{Benli_Ertan_17}. 

The initial condition of a source is defined by the initial disc mass ($\Md$), the initial period ($P_0$), and the magnitude of the magnetic dipole field on the pole of the star ($B_0$). In the model, different evolutionary avenues are produced by the differences in these initial conditions. We try to understand the characteristic differences in the initial conditions responsible for the emergence of different  neutron star populations (see e..g. \citealt{Benli_Ertan_16} for a comparison of AXP/SGRs and XDINs).    

In the model, a particular source at a given time could be in the accretion phase or in the propeller phase depending on the current mass-flow rate at the inner disc, period and dipole field strength of the source. In the propeller phase, the X-ray luminosity, $\Lx$, is produced mainly by the intrinsic cooling of the star, while in the accretion phase both the cooling and the mass accretion on to the star have contribution to $\Lx$. In the accretion phase,  the accretion luminosity usually dominates the intrinsic cooling of the star. We assume that the pulsed radio emission is allowed only in the propeller phase, since the mass-flow on to the star is likely to hinder the radio emission. The accretion luminosity equals to $G M \Mdot/R$ where $\Mdot$ is the rate of  accretion on to the star,  $G$ is the gravitational constant, $M$ and $R$ are the mass and the radius of the neutron star. For the cooling luminosity, we use the theoretical cooling curve estimated for neutron stars with conventional magnetic dipole fields \citep{Page_etal_04, Page_etal_06}. 

Part of the X-rays emitted by the neutron stars illuminates the disc surfaces. This X-ray irradiation flux together with viscous dissipation heats up the disc. Outside a few $10^9$ cm, depending on the irradiation efficiency, the irradiation flux is the dominant source of heating. The dynamical outer radius of the disc corresponds to the radius at which the effective temperature is currently equal to the critical temperature $\Tp$ below which the disc becomes viscously inactive. The irradiation flux at a radius $r$ can be written in terms of the the X-ray luminosity  as $\Firr = 1.2~C \Lx / (\pi r^2)$ \citep{Fukue_92}. The irradiation efficiency parameter $C$ depends on the geometry and the albedo of the disc surfaces. The properties of AXP/SGRs, XDINs, and HBRPs with known braking indices can be obtained with $\Tp \sim 100$ K and $C = (1 -7) \times 10^{-4}$, the same range as estimated for soft X-ray transients (see e.g. \citealt{Dubus_etal_01}). Following the results obtained from the analyses of X-ray enhancements light curves of AXP/SGRs \citep{Caliskan_Ertan_12}, we have calculated the kinematic viscosity with $\alpha = 0.045$ in the earlier applications of the model to different neutron star systems.  We use these main disc parameters in the present work as well.

We solve the disc diffusion equation for an extended, geometrically thin standard disc (see e.g. \citealt{Frank_etal_02}). The evolution of the disc also determines the rotational evolution of the star through interaction between the inner disc and the dipole field of the star. Details of the torque model are described in \citealt{Ertan_Erkut_08}. 

The condition for the accretion-propeller transition is not well known. In our model, we assume that the system is in the propeller phase when the calculated conventional \Alfven~radius $\rA \simeq (G M)^{-1/7}~\mu^{4/7} \Mdotin^{-2/7}$ is greater than the light cylinder radius $\rlc = c/\Omegastar$, where $\Omegastar$ and $\mu$ are the angular frequency and the magnetic dipole moment of the star, $c$ is the speed of light and $\Mdotin$ is the disc mass-flow rate at the inner disc.

\section{The Observational Properties of High--B Radio Pulsars}
\label{sec3}

{\bf PSR B1845--19 }~was discovered in the Molonglo Pulsar Survey with $P = 4.31$ s \citep{Manchester_etal_78} and $\Pdot = 2.33 \times 10^{-14}$ s s$^{-1}$. Its spin-down power $\Edot = 1.1 \times 10^{31}$~erg~s$^{-1}$, characteristic age  $\tau = P/{2 \Pdot} \simeq 2.9 \times 10^6$ yr and the field strength on the pole, inferred from the dipole torque formula, $\Bd = 2 \times 10^{13}$~G. A $3\sigma$ upper limit to the bolometric X-ray luminosity, $\Lx < 6.8 \times 10^{31}$~erg~s$^{-1}$, was obtained with an estimated distance $d = 0.75$~kpc \citep{Olausen_etal_13}.

{\bf PSR J1001--5939} was discovered in the Parkes Multibeam Pulsar Survey with $P = 7.73$~s, $\Pdot = 5.99 \times 10^{-14}$~s~s$^{-1}$ \citep{Lorimer_etal_06}, which give $\Edot = 5.1 \times 10^{30}$~erg~s$^{-1}$, $\tau \sim 2 \times 10^6$~yr and $\Bd = 4.3 \times 10^{13}$~G. A $3\sigma$ upper limit to the X-ray luminosity of PSR J1001--5939, from XMM-Newton and Chandra observations, is $\simeq 1.26 \times 10^{31}$~erg~s$^{-1}$ for $d = 2.7$~kpc \citep{Olausen_etal_13}.

{\bf PSR J1847--0130}~was discovered in the Parkes Multibeam Pulsar Survey with a period $P = 6.7$ s and period derivative $\Pdot = 1.3 \times 10^{-12}$ s s$^{-1}$ \citep{Hobbs_etal_04, Mclaughlin_etal_03}, which corresponds to $\Edot = 1.7  \times 10^{32}$ erg s$^{-1}$, $\tau \sim 8 \times 10^4$~yr and $\Bd = 1.9 \times 10^{14}$~G. The upper limit to the $0.5-10$ keV X-ray luminosity is $\sim 3.4 \times 10^{34}$ erg s$^{-1}$ for $d = 8.4$~kpc \citep{Mclaughlin_etal_03, Ng_Kaspi_11}.

{\bf PSR J1718--3718} has $P = 3.38$ s and $\Pdot = 1.5 \times 10^{-12}$ s s$^{-1}$ \citep{Hobbs_etal_04, Zhu_etal_11}, which give $\Edot = 1.7 \times 10^{33}$ erg s$^{-1}$, $\tau \sim 3.4 \times 10^4$ yr and $\Bd = 1.4 \times 10^{14}$~G. Recent \textit{Chandra} observations showed that the source also emits pulsed X-rays with the same period as the radio pulses \citep{Zhu_etal_11}. The persistent X-ray luminosity in the $0.5-10$ keV band is $\Lx = (0.14 - 2.60) \times 10^{33}$ erg s$^{-1}$ for $d = 4.5$ kpc \cite{Kaspi_Mclaughlin_05}.

{\bf PSR B1916+14} was discovered with $P = 1.18$ s and $\Pdot = 2.1 \times 10^{-13}$ s s$^{-1}$ \citep{Hulse_taylor_74}. The spin-down age is $\tau = 8.8 \times 10^4$ yr, rotational energy lost rate $\Edot = 5.1 \times 10^{33}$ erg s$^{-1}$ and the field strength $\Bd = 3,2 \times 10^{13}$~G . The distance to the source was estimated to be $d = 2.1\pm0.3$ kpc from the dispersion measure of the pulsar (see e.g. \citealt{Ng_Kaspi_11}). The bolometric luminosities from the blackbody and the power-law fits to the X-ray spectrum, are $ \sim 3 \times 10^{31}$ and $ \sim 2 \times 10^{31}$ erg s$^{-1}$, respectively \citep{Zhu_etal_09}. Considering the uncertainties in the estimated temperature of the source, we adopt  $\Lx \sim 10^{31}-10^{32}$~erg~s$^{-1}$.

\section{Results and Conclusions }
\label{sec4}

We have investigated the long-term evolutionary paths in the fallback disc model to determine the initial conditions ($B_0$, $P_0$, $\Md$) that can lead to the observed properties of the five HBRPs. Radio pulsars with $\Bd$ values greater than the quantum critical field $4.4 \times 10^{13}$~G are usually called HBRPs. In the fallback disc model, the $B_0$ values for different neutron star populations are estimated to be less than $10^{13}$~G, much lower than $\Bd$ values deduced from the observed rotational properties. That is, in this model, neutron stars with $\Bd \gtrsim 10^{13}$ G are good candidates for being among the systems evolving with  fallback discs. In this work, we select the sources with measured $P$, $\Pdot$ and $\Lx$ values, and with $\Bd > 2 \times 10^{13}$~G.

\begin{figure}
\centering
\includegraphics[width=0.7\columnwidth,angle=0]{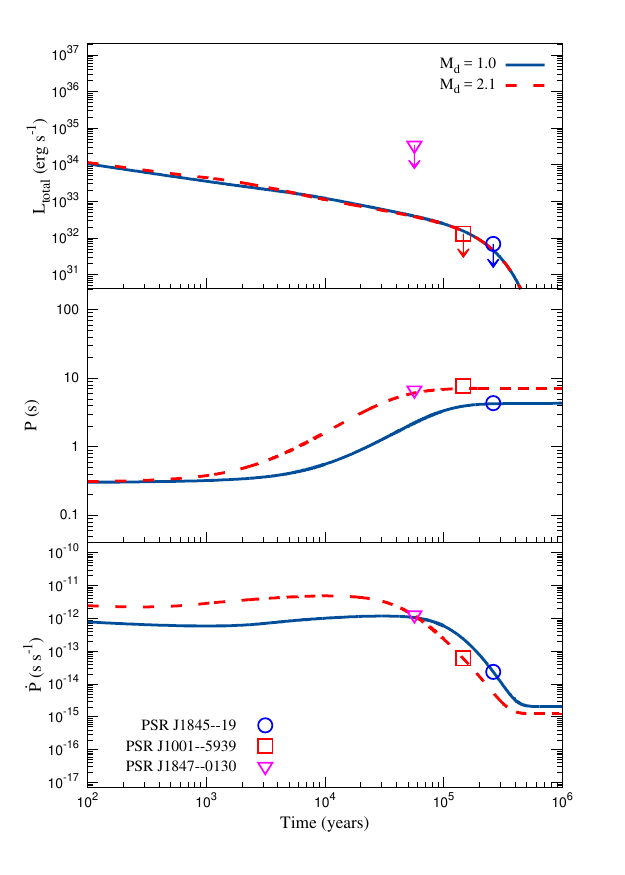}
\caption{Illustrative model curves for PSR B1845--19, PSR J1001--5939 and PSR J1847--0130. The data points with arrows in the top panel show the observed $\Lx$ upper limits. The data points in the middle and the bottom panels show the observed $P$ and $\Pdot$ values. Both model curves, are obtained with $B_0 = 6 \times 10^{12}$~G. The $\Md$ values (in units of $10^{-6}~\Msun$) are given in the top panel. The model parameters are given in Table \ref{tab:properties}. } 
\label{fig:3in1}
\end{figure}

\begin{figure}
\centering
\includegraphics[width=0.7\columnwidth,angle=0]{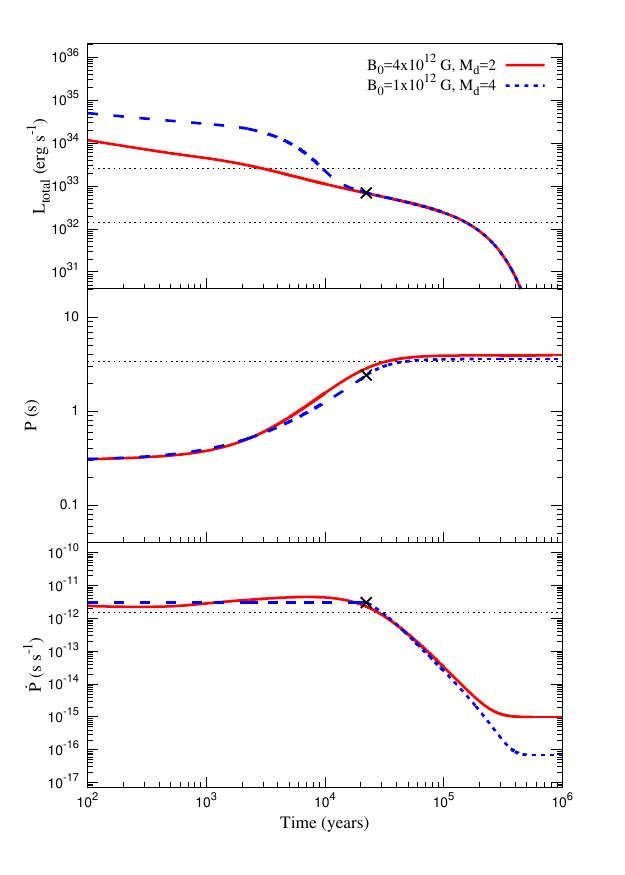}
\caption{Model curves for  PSR J1718--3718. The dashed (blue) curve represents the evolution of a source that is initially in the accretion phase (long-dashed curves), and makes a transition to the propeller phase (short-dashed curves) at $t \sim 2 \times 10^{4}$ yr (marked by cross signs on the model curves). For the evolution shown by the solid (red) curve, the source always remains in the propeller phase. The $\Md$ (in units of $10^{-6}~\Msun$) and $B_0$ values  are given in the top panel (see Table \ref{tab:properties}~for the parameters). Two horizontal dashed lines in the top panel show the observed $\Lx$ interval. The dashed lines in the middle and bottom panels show the observed $P$ and $\Pdot$ values for the sources.    }
\label{fig:1718}
\end{figure}

\begin{figure}
\centering
\includegraphics[width=0.7\columnwidth,angle=0]{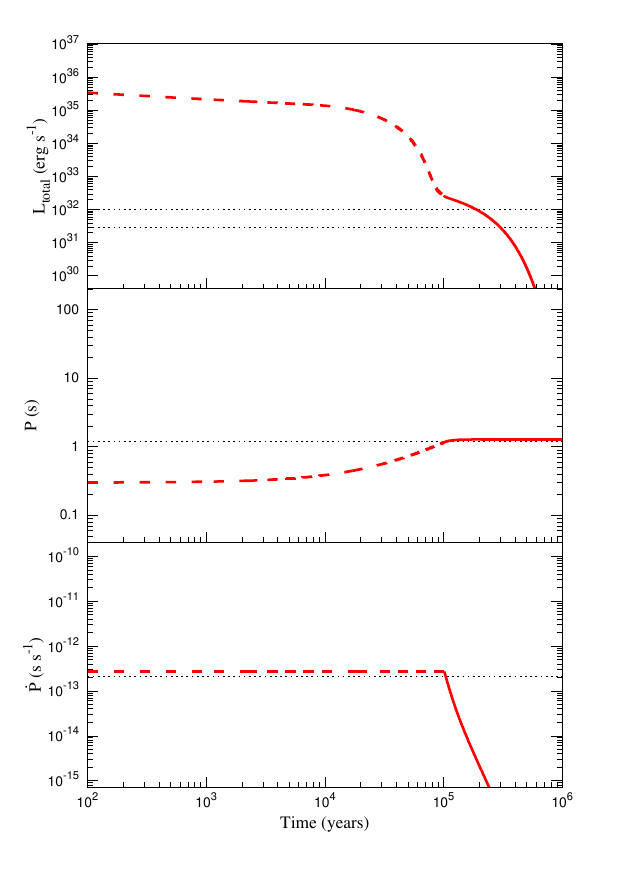}
\caption{An illustrative model curve that could represent the evolution of PSR B1916+14 with $B_0 = 3 \times 10^{11}$~G and $\Md \simeq 10^{-5}~\Msun$. The source properties are acquired shortly after the termination of the accretion phase at $t \sim 10^5$ yr. The accretion and propeller phases are represented by dashed and solid curves, respectively. The model parameters are given in Table \ref{tab:properties}. }
\label{fig:B1916}
\end{figure}

It is seen in Figs. \ref{fig:3in1}-\ref{fig:B1916} that our model can reproduce the periods and the period derivatives consistently with their X-ray luminosities for two sources and with the luminosity upper limits for three sources. 
The model parameters  are given in Table 1. We use the same basic disc parameters ($\alpha$, $\Tp$, $C$) that were also used in the earlier applications of the same model to AXP/SGRs, XDINs, and the 3 HBRPs with known braking indices.  The results of these analyses show that the individual source properties ($\Lx, P, \Pdot$) of all these neutron star populations can be reproduced in a single model without requiring any additional assumption for a particular source or population. 

The model curves given in Figs. \ref{fig:3in1}-\ref{fig:B1916} represents the evolutions of the sources that are currently in the propeller phase. Unlike for AXP/SGRs and XDINs, the model does not constrain the field strength for HBRPs. Using different  $P_0$ and  $\Md$ values similar results are obtained with a large range of  $B_0$ from $\sim 10^{12}$~G to the $\Bd$ values inferred with the dipole torque assumption. The main restriction on $B_0$ is imposed by the radio pulsar property of these systems, which allows $B_0$ values beyond a minimum critical value required for pulsed radio emission. This restriction is important especially for the sources with relatively long periods which need relatively strong fields for the radio emission. Furthermore, any solutions that produce the source properties in the accretion phase are not acceptable either, because the mass-flow on to the neutron star is expected to switch off the radio emission.    

For the three sources with model curves given in Fig. \ref{fig:3in1}, reasonable solutions imply that these systems have always evolved in the propeller phase as radio pulsars. It is seen in Fig. \ref{fig:3in1} that they have very similar evolutionary curves that are produced with $B_0 = 6 \times 10^{12}$~G. There are initial conditions leading to the properties of these sources with $B_0 \sim 10^{12}$~G, which are not compatible with the radio pulsar properties of these HBRPs.   
 
For PSR J1718--3718 (Fig. \ref{fig:1718}), there are two different evolutionary avenues that could represent the long-term evolution of the source. For the illustrative model with $B_0 = 4 \times 10^{12}$ G, the source always remains in the propeller phase, while for $B_0 = 1 \times 10^{12}$ G, the system is initially in the accretion phase, and makes a transition to the propeller phase at $t \sim 2 \times 10^4$ yr. For this source, these field strengths are sufficient for pulsed radio emission. Another source that was likely to be initially in the accretion phase is PSR B1916+14. The evolutionary curve in Fig. \ref{fig:B1916} indicates that this source entered the propeller phase a few 100 yr ago at $t \sim 10^5$ yr. For the illustrative models  with $B_0 = 6 \times 10^{12}$ G for  PSR B1845--19  and PSR J1001--5939, it is seen in Fig. 1 that the source properties are reproduced at a few times $10^{5}$ yr which are about an order of magnitude smaller than the characteristic ages of the sources. Because these model sources have evolved in their history with period derivatives larger than the values measured at present. In the future, the sharp decrease in their $\Pdot$ is likely to continue until the dipole torques dominate the disc torques, which is expected to happen when  $\Pdot$ decrease to a level of $\sim 10^{-15}$ s s$^{-1}$ depending on the actual strengths of their dipole fields. 

It is not possible to interpret the measured braking indices of these sources in a model due to timing noise and/or glitch effects \citep{Manchester_Hobbs_11, Hobbs_etal_10}. In our model, the sources in the late propeller phase of evolution could have rather high braking indices. For the illustrative models given in Figs. \ref{fig:3in1}-\ref{fig:B1916}, we estimate $n = 106$ for PSR B1845--19 and $n =7 - 8$ for PSR J1718--3718. For PSR B1916+14, $n$ increases rapidly from $ \sim 15$ at $t = 1.02 \times 10^5$ yr to $\sim 45$ at $t = 1.20 \times 10^5$ yr.

Our results together with those of earlier analyses of AXP/SGRs, XDINs, and HBRPs show that there are rather interesting evolutionary connections between these systems. In particular, the evolutionary path shown by the dashed curve in Fig. \ref{fig:1718} represent the evolution of an AXP to the HBRP phase, while the transition of an HBRP to the AXP phase is also possible (see e.g. \citealt{Caliskan_etal_13}). The details of these evolutionary connections between the neutron star populations will be studied in an independent work.        

\begin{table*}
\caption{Model paremeters and the properties of high--B radio pulsars (see Section \ref{sec3}~for the references). We take the viscosity parameter $\alpha = 0.045$ and the critical temperature $\Tp = 100$~K for all our model calculations.}
\label{tab:properties}
\bigskip
\centering
\footnotesize
\begin{tabular}{l|cccc|ccr}
\cline{1-8}
&\multicolumn{4}{c}{Observational Properties}  & \multicolumn{3}{c}{Model Parameters} \\ 
  		Name						 & \begin{tabular}{@{}c@{}}$P$\\(s)\end{tabular} &\begin{tabular}{@{}c@{}} $\Pdot$\\ (s s$^{-1}$)\end{tabular}       & \begin{tabular}{@{}c@{}}$\Lx$\\ (erg s$^{-1}$)\end{tabular}    	& \begin{tabular}{@{}c@{}}$d$\\ (kpc)\end{tabular} &  $C$  						&\begin{tabular}{@{}c@{}}$\Md$ \\ ($10^{-6}~\Msun$)\end{tabular}  & \begin{tabular}{@{}c@{}}$P_0$\\ (ms)\end{tabular}\\
\hline 
PSR B1845--19  		  & 4.31   & $2.33 \times 10^{-14}$   & $< 6.8 \times 10^{31}$    &0.75			& $7 \times 10^{-4}$ 	& 1.0 									&  300\\
PSR J1001--5939 	  & 7.73   & $5.99 \times 10^{-14}$   & $< 1.3 \times 10^{32}$    & 2.7			 &$1 \times 10^{-4}$ 	& 2.1 									& 300\\
PSR J1847--0130  	  & 6.71   & $1.28 \times 10^{-12}$   & $< 3.4 \times 10^{34}$    &8.4 				&$1 \times 10^{-4}$ 	& 2.1 									&  300\\
PSR J1718--3718		  & 3.38   & $1.61 \times 10^{-12}$   & $0.14-2.6 \times 10^{33}$ & 4.5				&$1 \times 10^{-4}$		& $2.0~\&~4.0$ 				&  300\\
PSR B1916+14		  & 1.18   & $2.12 \times 10^{-13}$   & $1.1-2.3 \times 10^{31}$  &$2.1\pm0.3$ &$7 \times 10^{-4}$ 	& 10.0 								&  300\\
\hline
\end{tabular}
\end{table*}

\section*{Acknowledgements}

{We acknowledge research support from
T\"{U}B{\.I}TAK (The Scientific and Technological Research Council of
Turkey) through grant 116F336  and  from Sabanc\i\ University. 

\newpage

\bibliographystyle{elsarticle_harv}
\bibliography{benli}

\begin{thebibliography}{36}
\expandafter\ifx\csname natexlab\endcsname\relax\def\natexlab#1{#1}\fi
\expandafter\ifx\csname url\endcsname\relax
  \def\url#1{\texttt{#1}}\fi
\expandafter\ifx\csname urlprefix\endcsname\relax\def\urlprefix{URL }\fi

\bibitem[{{Alpar}(2001)}]{Alpar_01}
{Alpar}, M.~A., Jun. 2001. {On Young Neutron Stars as Propellers and Accretors
  with Conventional Magnetic Fields}. ApJ 554, 1245--1254.

\bibitem[{{Benli} and {Ertan}(2016)}]{Benli_Ertan_16}
{Benli}, O., {Ertan}, {\"U}., Apr. 2016. {Long-term evolution of anomalous
  X-ray pulsars and soft gamma repeaters}. MNRAS 457, 4114--4122.

\bibitem[{{Benli} and {Ertan}(2017)}]{Benli_Ertan_17}
{Benli}, O., {Ertan}, {\"U}., 2017. {On the evolution of high-B radio pulsars
  with measured braking indices.} Submitted to MNRAS.

\bibitem[{{\c{C}al{\i}\c{s}kan} and {Ertan}(2012)}]{Caliskan_Ertan_12}
{\c{C}al{\i}\c{s}kan}, {\c S}., {Ertan}, {\"U}., Oct. 2012. {On the X-Ray
  Outbursts of Transient Anomalous X-Ray Pulsars and Soft Gamma-Ray Repeaters}.
  ApJ 758, 98.

\bibitem[{{\c{C}al{\i}\c{s}kan} et~al.(2013){\c{C}al{\i}\c{s}kan}, {Ertan},
  {Alpar}, {Tr{\"u}mper}, and {Kylafis}}]{Caliskan_etal_13}
{\c{C}al{\i}\c{s}kan}, {\c S}., {Ertan}, {\"U}., {Alpar}, M.~A., {Tr{\"u}mper},
  J.~E., {Kylafis}, N.~D., May 2013. {On the evolution of the radio pulsar PSR
  J1734-3333}. MNRAS 431, 1136--1142.

\bibitem[{{Chatterjee} et~al.(2000){Chatterjee}, {Hernquist}, and
  {Narayan}}]{Chatterjee_etal_00}
{Chatterjee}, P., {Hernquist}, L., {Narayan}, R., May 2000. {An Accretion Model
  for Anomalous X-Ray Pulsars}. ApJ 534, 373--379.

\bibitem[{{Chevalier}(1989)}]{Chevalier_89}
{Chevalier}, R.~A., Nov. 1989. {Neutron star accretion in a supernova}. ApJ
  346, 847--859.

\bibitem[{{Dubus} et~al.(2001){Dubus}, {Hameury}, and {Lasota}}]{Dubus_etal_01}
{Dubus}, G., {Hameury}, J.-M., {Lasota}, J.-P., Jul. 2001. {The disc
  instability model for X-ray transients: Evidence for truncation and
  irradiation}. \aap 373, 251--271.

\bibitem[{{Ertan} et~al.(2014){Ertan}, {\c{C}al{\i}\c{s}kan}, {Benli}, and
  {Alpar}}]{Ertan_etal_14}
{Ertan}, {\"U}., {\c{C}al{\i}\c{s}kan}, {\c S}., {Benli}, O., {Alpar}, M.~A.,
  Oct. 2014. {Long-term evolution of dim isolated neutron stars}. MNRAS 444,
  1559--1565.

\bibitem[{{Ertan} and {Erkut}(2008)}]{Ertan_Erkut_08}
{Ertan}, {\"U}., {Erkut}, M.~H., Feb. 2008. {On the X-Ray Light Curve,
  Pulsed-Radio Emission, and Spin Frequency Evolution of the Transient
  Anomalous X-Ray Pulsar XTE J1810-197 during Its X-Ray Outburst}. ApJ 673,
  1062--1066.

\bibitem[{{Frank} et~al.(2002){Frank}, {King}, and {Raine}}]{Frank_etal_02}
{Frank}, J., {King}, A., {Raine}, D.~J., Jan. 2002. {Accretion Power in
  Astrophysics: Third Edition}.

\bibitem[{{Fukue}(1992)}]{Fukue_92}
{Fukue}, J., Dec. 1992. {Self-irradiated accretion disks}. \pasj 44, 663--667.

\bibitem[{{Gavriil} et~al.(2008){Gavriil}, {Gonzalez}, {Gotthelf}, {Kaspi},
  {Livingstone}, and {Woods}}]{Gavriil_etal_08}
{Gavriil}, F.~P., {Gonzalez}, M.~E., {Gotthelf}, E.~V., {Kaspi}, V.~M.,
  {Livingstone}, M.~A., {Woods}, P.~M., Mar. 2008. {Magnetar-Like Emission from
  the Young Pulsar in Kes 75}. Science 319, 1802.

\bibitem[{{Hobbs} et~al.(2004){Hobbs}, {Faulkner}, {Stairs}, {Camilo},
  {Manchester}, {Lyne}, {Kramer}, {D'Amico}, {Kaspi}, {Possenti}, {McLaughlin},
  {Lorimer}, {Burgay}, {Joshi}, and {Crawford}}]{Hobbs_etal_04}
{Hobbs}, G., {Faulkner}, A., {Stairs}, I.~H., {Camilo}, F., {Manchester},
  R.~N., {Lyne}, A.~G., {Kramer}, M., {D'Amico}, N., {Kaspi}, V.~M.,
  {Possenti}, A., {McLaughlin}, M.~A., {Lorimer}, D.~R., {Burgay}, M., {Joshi},
  B.~C., {Crawford}, F., Aug. 2004. {The Parkes multibeam pulsar survey - IV.
  Discovery of 180 pulsars and parameters for 281 previously known pulsars}.
  \mnras 352, 1439--1472.

\bibitem[{{Hobbs} et~al.(2010){Hobbs}, {Lyne}, and {Kramer}}]{Hobbs_etal_10}
{Hobbs}, G., {Lyne}, A.~G., {Kramer}, M., Feb. 2010. {An analysis of the timing
  irregularities for 366 pulsars}. \mnras 402, 1027--1048.

\bibitem[{{Hulse} and {Taylor}(1974)}]{Hulse_taylor_74}
{Hulse}, R.~A., {Taylor}, J.~H., Jul. 1974. {A High-Sensitivity Pulsar Survey}.
  \apjl 191, L59.

\bibitem[{{Israel} et~al.(2007){Israel}, {Campana}, {Dall'Osso}, {Muno},
  {Cummings}, {Perna}, and {Stella}}]{Israel_etal_07}
{Israel}, G.~L., {Campana}, S., {Dall'Osso}, S., {Muno}, M.~P., {Cummings}, J.,
  {Perna}, R., {Stella}, L., Jul. 2007. {The Post-Burst Awakening of the
  Anomalous X-Ray Pulsar in Westerlund 1}. \apj 664, 448--457.

\bibitem[{{Kaspi}(2010)}]{Kaspi_10}
{Kaspi}, V.~M., Apr. 2010. {Grand unification of neutron stars}. Proceedings of
  the National Academy of Science 107, 7147--7152.

\bibitem[{{Kaspi} et~al.(2003){Kaspi}, {Gavriil}, {Woods}, {Jensen}, {Roberts},
  and {Chakrabarty}}]{Kaspi_etal_03}
{Kaspi}, V.~M., {Gavriil}, F.~P., {Woods}, P.~M., {Jensen}, J.~B., {Roberts},
  M.~S.~E., {Chakrabarty}, D., May 2003. {A Major Soft Gamma Repeater-like
  Outburst and Rotation Glitch in the No-longer-so-anomalous X-Ray Pulsar 1E
  2259+586}. ApJL 588, L93--L96.

\bibitem[{{Kaspi} and {McLaughlin}(2005)}]{Kaspi_Mclaughlin_05}
{Kaspi}, V.~M., {McLaughlin}, M.~A., Jan. 2005. {Chandra X-Ray Detection of the
  High Magnetic Field Radio Pulsar PSR J1718-3718}. \apjl 618, L41--L44.

\bibitem[{{Keane} and {Kramer}(2008)}]{Keane_Kramer_08}
{Keane}, E.~F., {Kramer}, M., Dec. 2008. {On the birthrates of Galactic neutron
  stars}. MNRAS 391, 2009--2016.

\bibitem[{{Lorimer} et~al.(2006){Lorimer}, {Faulkner}, {Lyne}, {Manchester},
  {Kramer}, {McLaughlin}, {Hobbs}, {Possenti}, {Stairs}, {Camilo}, {Burgay},
  {D'Amico}, {Corongiu}, and {Crawford}}]{Lorimer_etal_06}
{Lorimer}, D.~R., {Faulkner}, A.~J., {Lyne}, A.~G., {Manchester}, R.~N.,
  {Kramer}, M., {McLaughlin}, M.~A., {Hobbs}, G., {Possenti}, A., {Stairs},
  I.~H., {Camilo}, F., {Burgay}, M., {D'Amico}, N., {Corongiu}, A., {Crawford},
  F., Oct. 2006. {The Parkes Multibeam Pulsar Survey - VI. Discovery and timing
  of 142 pulsars and a Galactic population analysis}. \mnras 372, 777--800.

\bibitem[{{Manchester} and {Hobbs}(2011)}]{Manchester_Hobbs_11}
{Manchester}, R.~N., {Hobbs}, G., Aug. 2011. {A Giant Glitch in PSR
  J1718-3718}. \apjl 736, L31.

\bibitem[{{Manchester} et~al.(1978){Manchester}, {Lyne}, {Taylor}, {Durdin},
  {Large}, and {Little}}]{Manchester_etal_78}
{Manchester}, R.~N., {Lyne}, A.~G., {Taylor}, J.~H., {Durdin}, J.~M., {Large},
  M.~I., {Little}, A.~G., Nov. 1978. {The second Molonglo pulsar survey -
  discovery of 155 pulsars.} \mnras 185, 409--421.

\bibitem[{{McLaughlin} et~al.(2003){McLaughlin}, {Stairs}, {Kaspi}, {Lorimer},
  {Kramer}, {Lyne}, {Manchester}, {Camilo}, {Hobbs}, {Possenti}, {D'Amico}, and
  {Faulkner}}]{Mclaughlin_etal_03}
{McLaughlin}, M.~A., {Stairs}, I.~H., {Kaspi}, V.~M., {Lorimer}, D.~R.,
  {Kramer}, M., {Lyne}, A.~G., {Manchester}, R.~N., {Camilo}, F., {Hobbs}, G.,
  {Possenti}, A., {D'Amico}, N., {Faulkner}, A.~J., Jul. 2003. {PSR J1847-0130:
  A Radio Pulsar with Magnetar Spin Characteristics}. \apjl 591, L135--L138.

\bibitem[{{Michel}(1988)}]{Michel_88}
{Michel}, F.~C., Jun. 1988. {Neutron star disk formation from supernova
  fall-back and possible observational consequences}. Nature 333, 644.

\bibitem[{{Michel} and {Dessler}(1981)}]{Michel_Dessler_81}
{Michel}, F.~C., {Dessler}, A.~J., Dec. 1981. {Pulsar disk systems}. ApJ 251,
  654--664.

\bibitem[{{Ng} and {Kaspi}(2011)}]{Ng_Kaspi_11}
{Ng}, C.-Y., {Kaspi}, V.~M., Sep. 2011. {High Magnetic Field Rotation-powered
  Pulsars}. In: {G{\"o}{\u g}{\"u}{\c s}}, E., {Belloni}, T., {Ertan}, {\"U}.
  (Eds.), American Institute of Physics Conference Series. Vol. 1379 of
  American Institute of Physics Conference Series. pp. 60--69.

\bibitem[{{Olausen} et~al.(2013){Olausen}, {Zhu}, {Vogel}, {Kaspi}, {Lyne},
  {Espinoza}, {Stappers}, {Manchester}, and {McLaughlin}}]{Olausen_etal_13}
{Olausen}, S.~A., {Zhu}, W.~W., {Vogel}, J.~K., {Kaspi}, V.~M., {Lyne}, A.~G.,
  {Espinoza}, C.~M., {Stappers}, B.~W., {Manchester}, R.~N., {McLaughlin},
  M.~A., Feb. 2013. {X-Ray Observations of High-B Radio Pulsars}. \apj 764, 1.

\bibitem[{{Page} et~al.(2006){Page}, {Geppert}, and {Weber}}]{Page_etal_06}
{Page}, D., {Geppert}, U., {Weber}, F., Oct. 2006. {The cooling of compact
  stars}. Nuclear Physics A 777, 497--530.

\bibitem[{{Page} et~al.(2004){Page}, {Lattimer}, {Prakash}, and
  {Steiner}}]{Page_etal_04}
{Page}, D., {Lattimer}, J.~M., {Prakash}, M., {Steiner}, A.~W., Dec. 2004.
  {Minimal Cooling of Neutron Stars: A New Paradigm}. \apjs 155, 623--650.

\bibitem[{{Shakura} and {Sunyaev}(1973)}]{Shakura_Sunyaev_73}
{Shakura}, N.~I., {Sunyaev}, R.~A., 1973. {Black holes in binary systems.
  Observational appearance.} A\&A 24, 337--355.

\bibitem[{{Thompson} and {Duncan}(1995)}]{Thompson_Duncan_95}
{Thompson}, C., {Duncan}, R.~C., Jul. 1995. {The soft gamma repeaters as very
  strongly magnetized neutron stars - I. Radiative mechanism for outbursts}.
  MNRAS 275, 255--300.

\bibitem[{{Younes} et~al.(2016){Younes}, {Kouveliotou}, and
  {Roberts}}]{Younes_etal_16}
{Younes}, G., {Kouveliotou}, C., {Roberts}, O., 2016. {GBM observation of
  SGR-like burst from the direction of PSR 1119-6127.} GRB Coordinates Network
  19736.

\bibitem[{{Zhu} et~al.(2009){Zhu}, {Kaspi}, {Gonzalez}, and
  {Lyne}}]{Zhu_etal_09}
{Zhu}, W., {Kaspi}, V.~M., {Gonzalez}, M.~E., {Lyne}, A.~G., Oct. 2009.
  {Xmm-Newton X-Ray Detection of the High-Magnetic-Field Radio Pulsar PSR
  B1916+14}. \apj 704, 1321--1326.

\bibitem[{{Zhu} et~al.(2011){Zhu}, {Kaspi}, {McLaughlin}, {Pavlov}, {Ng},
  {Manchester}, {Gaensler}, and {Woods}}]{Zhu_etal_11}
{Zhu}, W.~W., {Kaspi}, V.~M., {McLaughlin}, M.~A., {Pavlov}, G.~G., {Ng},
  C.-Y., {Manchester}, R.~N., {Gaensler}, B.~M., {Woods}, P.~M., Jun. 2011.
  {Chandra Observations of the High-magnetic-field Radio Pulsar J1718-3718}.
  \apj 734, 44.

\end{thebibliography}

\end{document}